\documentclass[draftcls, onecolumn, journal, 12pt]{IEEEtran}

\usepackage{amsfonts}
\usepackage{amssymb}
\usepackage{graphicx}
\usepackage{hyperref, epstopdf}

\begin{document}

\title{New approximations for DQPSK transmission bit
error rate}

\author{Szil\'ard~Andr\'as, \'Arp\'ad~Baricz

\thanks{The research of \'A. Baricz was supported by the J\'anos Bolyai
Research Scholarship of the Hungarian Academy of Sciences and by the Romanian
National Council for Scientific Research in Education CNCSIS-UEFISCSU, project number PN-II-RU-PD\underline{ }388/2012. The authors are indebted to their friend Yin Sun for his useful comments.}
\thanks{S. Andr\'as is with Department of Applied Mathematics, Babe\c{s}-Bolyai
University, Cluj-Napoca 400084, Romania (e-mail:
andraszk@yahoo.com).}
\thanks{\'A. Baricz is with Department of Economics, Babe\c{s}-Bolyai
University, Cluj-Napoca 400591, Romania (e-mail:
bariczocsi@yahoo.com).} }

\maketitle

\begin{abstract}
In this correspondence our aim is to use some tight lower and upper bounds for the differential quaternary phase shift
keying transmission bit error rate in order to deduce accurate approximations for the bit error rate
by improving the known results in the literature. The computation of our new approximate expressions
are significantly simpler than that of the exact expression.
\end{abstract}


\section{Introduction}

Recently, Ferrari and Corazza
\cite{cofe} have studied the performance analysis of the
differential quaternary phase shift keying (DQPSK) transmission with
Gray coding over the additive white Gaussian noise (AWGN) channel. We note that
the bit error rate (BER) in the case of Gray coding can be written in terms of the Marcum $Q$-function
and the modified Bessel function of the first kind and zero order.
In \cite{cofe} the authors used some bounds for the Marcum $Q$-function
in order to propose some simple, but accurate approximations for the BER. By using some other bounds for the Marcum $Q$-function
(deduced by Wang and Wu \cite{wang}, and Baricz and Sun \cite{baricz}), very recently
Sun et al. \cite{sun} proposed another approximation for BER. In this correspondence
we make a contribution to the subject by using some new results on Marcum
$Q$-function: we deduce a new tight upper bound for the BER and combining this with
some existing tight bounds we construct some new accurate approximations. These results improve and complement the results of Ferrari and
Corazza \cite{cofe}, and also of Sun et al. \cite{sun}. In Section \ref{section2} we discuss the bounds for BER of DQPSK, in Section \ref{section3} we present five new approximations for BER and we compare them with the similar ones deduced from inequalities for the Marcum Q-function. Finally, in Section \ref{section4} the conclusion of the paper is given.

\section{Bounds for BER of DQPSK}\label{section2}

It is known that for the DQPSK transmission with Gray coding over an AWGN
channel the BER can expressed as follows \cite{pro}:
\begin{equation}\label{eq4}
\mbox{BER}\triangleq Q(a,b)-\frac{1}{2}I_0(ab)e^{-\frac{a^2+b^2}{2}},
\end{equation}
where $a=\sqrt{\gamma(2-\sqrt{2})},$
$b=\sqrt{\gamma(2+\sqrt{2})}$ and $\gamma$ is the bit
signal-to-noise ratio (SNR). Here $Q(a,b)$ stands for the Marcum
$Q$-function, defined by
$$Q(a,b)\triangleq \int_b^{\infty}xe^{-\frac{x^2+a^2}{2}}I_0(ax)dx,$$
where $b\geq 0,$ $a>0$ and $I_0$ is the modified Bessel
function of the first kind and zero order. Note that the Marcum $Q$-function play an
important role in the studies of digital communications over fading
channels \cite{Simon}.

Now, let us consider the complementary error function, defined by
$$\textrm{erfc}(x)\triangleq\frac{2}{\sqrt{\pi}}\int_x^{\infty}e^{-t^2}dt,$$
and for shortness let us introduce the notation
$$e(a,b)\triangleq\textrm{erfc}\left(\frac{b-a}{\sqrt{2}}\right)=\frac{2}{\sqrt{\pi}}\int_{\frac{b-a}{\sqrt{2}}}^{\infty}e^{-t^2}dt.$$
Recently, Ferrari and Corazza \cite{cofe} by using their bounds from \cite{corazza} (see also \cite{baricztran}) for the Marcum $Q$-function
\begin{equation}\label{eq1}Q(a,b)\geq\sqrt{\frac{\pi}{2}}\frac{bI_0(ab)}{e^{ab}}e(a,b),\end{equation}
\begin{equation}\label{eq2}Q(a,b)\leq\frac{I_0(ab)}{e^{ab}}\left[e^{-\frac{(b-a)^2}{2}}+a\sqrt{\frac{\pi}{2}}
e(a,b)\right],\end{equation}
where $b\geq a>0,$
deduced the following bounds for BER of DQPSK
with Gray coding:
$$\textrm{BER}>I_0(ab)\left[\sqrt{\frac{\pi}{2}}\frac{b}{e^{ab}}e(a,b)-
\frac{1}{2}e^{-\frac{a^2+b^2}{2}}\right]\triangleq\mbox{L}_1,$$
$$\textrm{BER}<I_0(ab)\left[\sqrt{\frac{\pi}{2}}\frac{a}{e^{ab}}e(a,b)
+\frac{1}{2}e^{-\frac{a^2+b^2}{2}}\right]\triangleq\mbox{U}_1.$$
It is important to mention here that motivated by the above results of Ferrari and Corazza,
very recently Wang and Wu \cite{wang} proved that
$$Q(a,b)\geq
\sqrt{\frac{\pi}{2}}\frac{bI_0(ab)}{e^{ab}-e^{-ab}}E(a,b),$$
while Baricz and Sun \cite{baricz} proved that
$$Q(a,b)\leq\frac{I_0(ab)}{e^{ab}+e^{-ab}}\left[e^{-\frac{(b-a)^2}{2}}+e^{-\frac{(b+a)^2}{2}}+
a\sqrt{\frac{\pi}{2}}E(a,b)\right],$$
where in both of the inequalities $b\geq a>0$ and
$$E(a,b)\triangleq e(a,b)-e(-a,b)=\frac{2}{\sqrt{\pi}}\int_{\frac{b-a}{\sqrt{2}}}^{\frac{b+a}{\sqrt{2}}}e^{-t^2}dt.$$
Now, by using the above lower and upper bounds for the Marcum $Q$-function, Sun et al. \cite{sun} deduced the following:
$$\mbox{BER}>I_0(ab)\left[\sqrt{\frac{\pi}{2}}\frac{b}{e^{ab}-e^{-ab}}E(a,b)-\frac{1}{2}e^{-\frac{a^2+b^2}{2}}\right]\triangleq\mbox{L}_2,$$
$$\mbox{BER}<I_0(ab)\left[\sqrt{\frac{\pi}{2}}\frac{a}{e^{ab}+e^{-ab}}E(a,b)+\frac{1}{2}e^{-\frac{a^2+b^2}{2}}\right]\triangleq\mbox{U}_2.$$
We note that the lower and upper bounds for the Marcum $Q$-function of Wang and Wu \cite{wang} and of Baricz and Sun \cite{baricz}, mentioned above,
are tighter than (\ref{eq1}) and (\ref{eq2}), as it was pointed out in \cite{baricz,wang}. This in turn implies that the bounds $\mbox{L}_2$ and $\mbox{U}_2$ are tighter than the bounds $\mbox{L}_1$ and $\mbox{U}_1,$ that is, we have $\mbox{L}_2>\mbox{L}_1$ and $\mbox{U}_2<\mbox{U}_1.$

\begin{figure}
\centering
\includegraphics[width=0.7\textwidth]{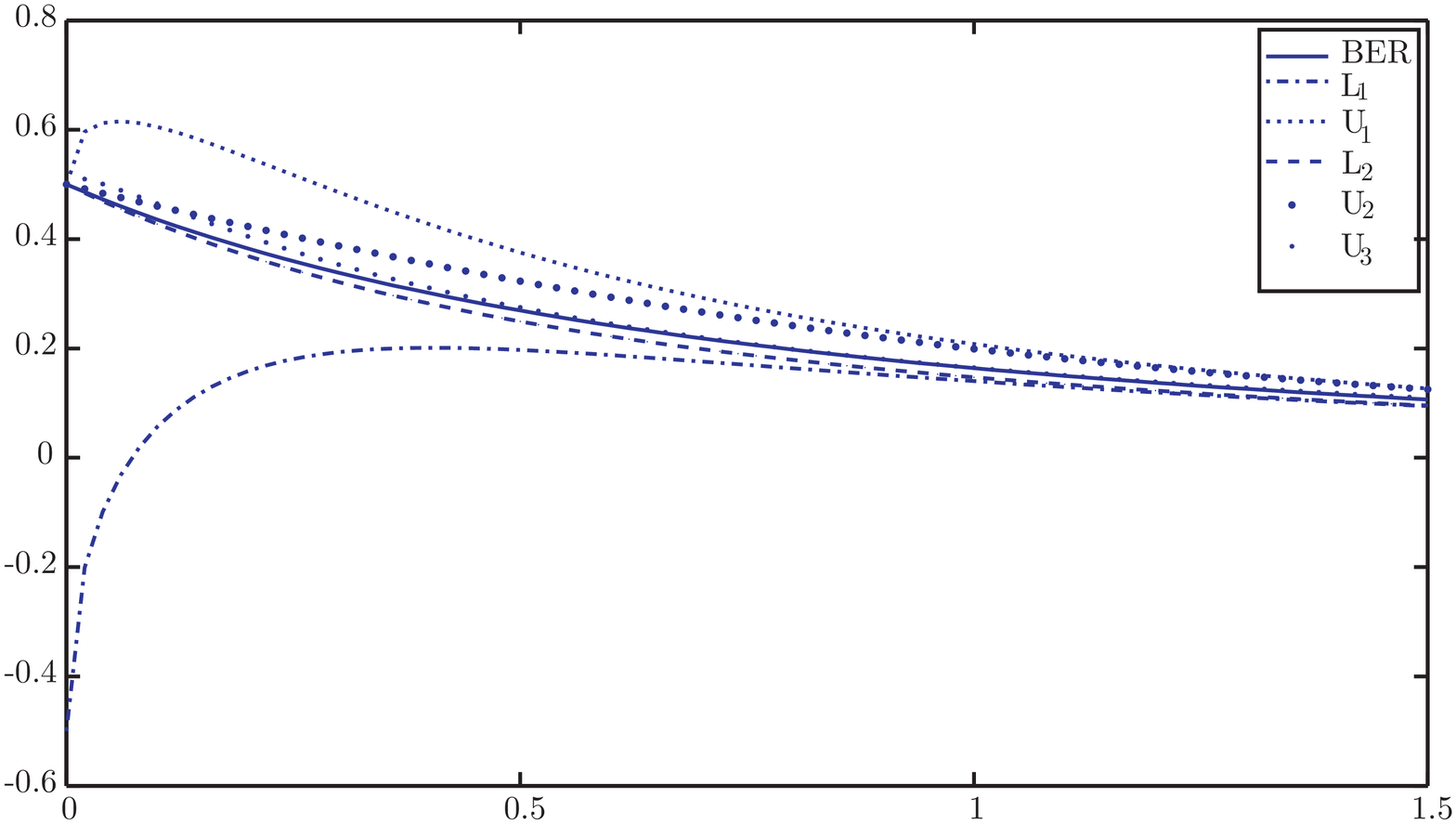}
\caption{BER with lower and upper bounds}\label{figure11}
\end{figure}

Now, let us consider the following inequality
$$Q(a,b)\leq \frac{I_0(ab)}{e^{ab}+\lambda_0}\left[a\sqrt{\frac{\pi}{2}}e(a,b)+(e^{ab}+\lambda_0)e^{-\frac{a^2+b^2}{2}}\right],$$
which holds for all $b\geq a>0$ and it was deduced by Baricz \cite{bariczjmaa}. Here $\lambda_0=e^{\rho_0}(I_0(\rho_0)/I_1(\rho_0)-1)\simeq3.03442206626763,$ $\rho_0\simeq1.54512596391949$ is the unique simple positive root of the
equation $(x+1)I_1(x) = xI_0(x),$ and $I_1$ stands for the modified Bessel function of order $1.$ We note that in the above inequality the constant $\lambda_0$
is best possible, that is, cannot be replaced by any larger constant. Since the above inequality is an improvement of (\ref{eq2}), its equivalent form
$$\mbox{BER}<I_0(ab)\left[\sqrt{\frac{\pi}{2}}\frac{a}{e^{ab}+\lambda_0}e(a,b)
+\frac{1}{2}e^{-\frac{a^2+b^2}{2}}\right]\triangleq\mbox{U}_3$$
clearly provides a better upper bound than (\ref{eq2}), that is, we have $\mbox{U}_3<\mbox{U}_1.$ The lower and upper bounds $\mbox{L}_1,$ $\mbox{L}_2,$ $\mbox{U}_1,$ $\mbox{U}_2$ and $\mbox{U}_3$ are shown in Fig. \ref{figure11} together with the exact BER as functions of the SNR on the interval $(0,1.5).$ Surprisingly, all of these bounds are extremely tight.

\section{Approximations for BER of DQPSK}\label{section3}

Observe that there is an interesting formal symmetry between the
bounds $\mbox{L}_1$ and $\mbox{U}_1,$ as well as between the bounds $\mbox{L}_2$ and $\mbox{U}_2.$
Based on the tight bounds $\mbox{L}_1$ and $\mbox{U}_1$, and
in view of the formal symmetry, Ferrari and Corazza
\cite{cofe} pointed out that it is natural to approximate the
BER simply by considering the arithmetic mean of the quantities
$\mbox{L}_1$ and $\mbox{U}_1.$
In this spirit, they derived the following approximate expression
for BER:
$$\mbox{BER}_1\triangleq\mathcal{A}\left(\mbox{L}_1,\mbox{U}_1;0.5\right),$$
i.e.
$$\mbox{BER}_1=
\sqrt{\frac{\pi}{8}}\frac{a+b}{e^{ab}}I_0(ab)e(a,b),$$
where $\mathcal{A}(x,y;\omega)=\omega x+(1-\omega)y$ is the weighted arithmetic mean of
$x$ and $y$ with weights $\omega,1-\omega>0.$ By using the same idea, it is natural to propose the
following new approximate expressions for BER:
$$\mbox{BER}_2\triangleq\mathcal{A}\left(\mbox{L}_2,\mbox{U}_2;0.5\right)$$
and
$$\mbox{BER}_3\triangleq\mathcal{A}\left(\mbox{L}_2,\mbox{U}_3;0.5\right),$$
that is,
$$\mbox{BER}_2=
\sqrt{\frac{\pi}{8}}\frac{(a+b)e^{ab}-(a-b)e^{-ab}}{e^{2ab}-e^{-2ab}}I_0(ab)E(a,b)$$
and
$$\mbox{BER}_3=
\sqrt{\frac{\pi}{8}}I_0(ab)\left[\frac{bE(a,b)}{e^{ab}-e^{-ab}}+\frac{ae(a,b)}{e^{ab}+\lambda_0}\right].$$
We note that these approximations are better than $\mbox{BER}_1.$  A
similar approximation to that of $\mbox{BER}_2$ was proposed recently
by Sun et al. \cite{sun} in the form
$$\mbox{BER}_4=\frac{e^{-\frac{(b+a)^2}{2}}}{\sqrt{8\pi ab}}+\frac{1}{4}\left(\sqrt{\frac{a}{b}}+\sqrt{\frac{b}{a}}\right)E(a,b),$$
however, this is better than $\mbox{BER}_1$ only for $\gamma\in(0.9,1.4).$ Motivated by the above results, in this correspondence our aim is to propose the following approximations based on better weight functions $\omega_i,$ $ i\in\{5,6,7\} :$
$$\mbox{BER}_5\triangleq\mathcal{A}\left(\mbox{L}_1,\mbox{U}_1;\omega_5\right),$$
$$\mbox{BER}_6\triangleq\mathcal{A}\left(\mbox{L}_2,\mbox{U}_2;\omega_6\right),$$
$$\mbox{BER}_7\triangleq\mathcal{A}\left(\mbox{L}_2,\mbox{U}_3;\omega_7\right),$$
where
$$\omega_5(\gamma)=\left \{\begin{array}{cl} 0.65\sqrt[4]{\gamma},&\mbox{if}\ \gamma<1\\
0.5+1.1\cdot\frac{ e^{\frac{-\pi}{2\sqrt{\gamma}}}} {\gamma^{1.5}}\cdot \sqrt{\frac 12},& \mbox{if}\ 1\leq \gamma
 \end{array} \right .,$$
$$\omega_6(\gamma)=\left \{\begin{array}{cl} e^{-\frac{\gamma^2}{2.9}}\cdot 0.25+0.5, & \mbox{if}\ 0\leq \gamma < 1\\
\frac{e^{-\frac{1}{2\gamma+1}}}{(\gamma+0.5)^{1.5}}\cdot \sqrt{\frac{1}{2\pi}}\cdot 1.15+0.5, &  \mbox{if}\ 1\leq \gamma < 5\\
\frac{1}{\pi} \left(1+\gamma \right)^{-1}\cdot 0.65+0.5, &\mbox{if}\ 5\leq \gamma \end{array} \right.,$$
$$\omega_7(\gamma)=\left \{\begin{array}{cl}  (1-\gamma)^2\cdot 0.95, & \mbox{if}\ 0\leq \gamma < 1\\
0.5-1.4\cdot e^{-\gamma^{1.2}}+0.02,& \mbox{if}\ 1\leq \gamma \leq 8 \\
\frac{1}{5.2\gamma}+0.5,&\mbox{if}\ 8< \gamma
  \end{array} \right ..$$
These approximations can be rewritten as follows
$$\mbox{BER}_5=\omega_5(\gamma)\mbox{L}_1+(1-\omega_5(\gamma))\mbox{U}_1,$$
$$\mbox{BER}_6=\omega_6(\gamma)\mbox{L}_2+(1-\omega_6(\gamma))\mbox{U}_2,$$
$$\mbox{BER}_7=\omega_7(\gamma)\mbox{L}_2+(1-\omega_7(\gamma))\mbox{U}_3.$$
It is important to note here that the computations of approximate
BER expressions $\mbox{BER}_i,$ where $i\in\{1,2,\dots,7\},$
completely avoid the computation of the Marcum $Q$-function. Tables
\ref{table1} and \ref{table2} contain an
estimate of BER by using the Matlab function \verb"marcumq" (BER)
and the estimates based upon $\mbox{BER}_i,$ where
$i\in\{1,\dots,7\}.$ As we can see the approximations
$\mbox{BER}_i,$ where $i\in\{5,6,7\},$ are better than the
approximations $\mbox{BER}_i,$ where $i\in\{1,2,3,4\}.$ Moreover,
the novel approximate expressions $\mbox{BER}_i,$ where
$i\in\{5,6,7\},$ are consistent with the results of Weinberg
\cite{weinberg}, obtained by using of Monte-Carlo estimators. We
note that in our further calculations we considered the values
obtained by the \verb"marcumq" function as the ``exact'' values for
BER.

\begin{figure}
\centering
\includegraphics[width=0.8\textwidth]{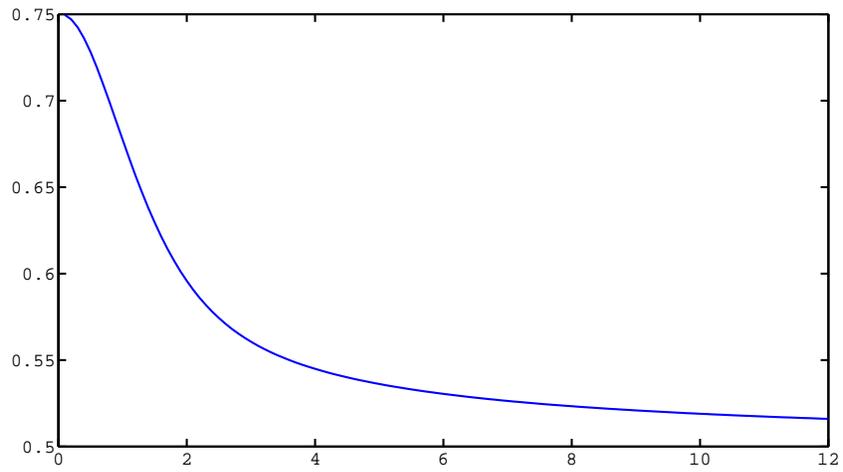}
\caption{The graph of the weight function $\omega_6$ }\label{figure7}
\end{figure}

We give a hint about the construction of the weight functions. By using the values of BER, $\mbox{L}_1,$ $\mbox{L}_2,$ $\mbox{U}_1,$ $\mbox{U}_2$ and $\mbox{U}_3$ first we calculated numerically the values of the weight functions. The representation of the initial numerical values for $\omega_6$ can be seen in Fig. \ref{figure7}. This looks like the probability density function of the normal distribution but having a heavy tail, so we started to search for expression involved in probability density functions with heavy tails (eg. Student, L\'evy,  $\alpha-$stable). By using the same idea the parameters in the expression of the functions $\omega_i,$ $i\in\{5,6,7\},$ were obtained by numerical experimentations.

\begin{figure}
\centering
\includegraphics[width=0.7\textwidth]{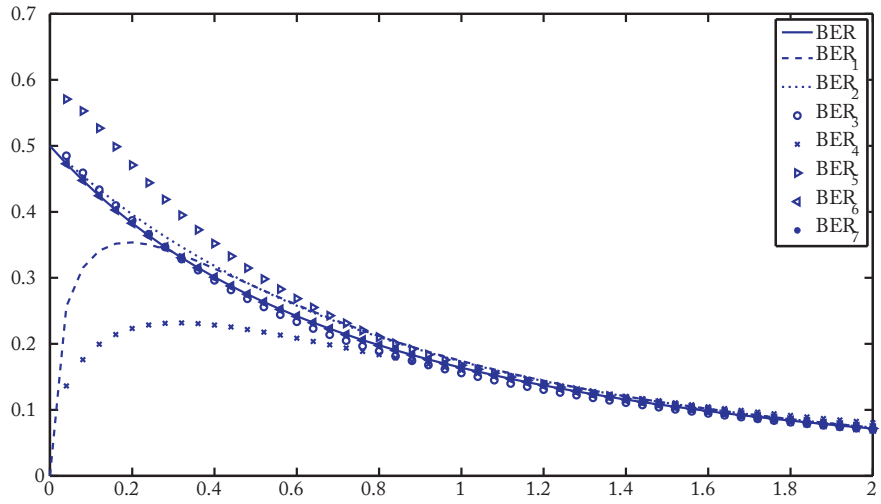}
\caption{Approximations for BER on $[0,2]$}\label{figure2}
\end{figure}

\begin{figure}
\centering
\includegraphics[width=0.7\textwidth]{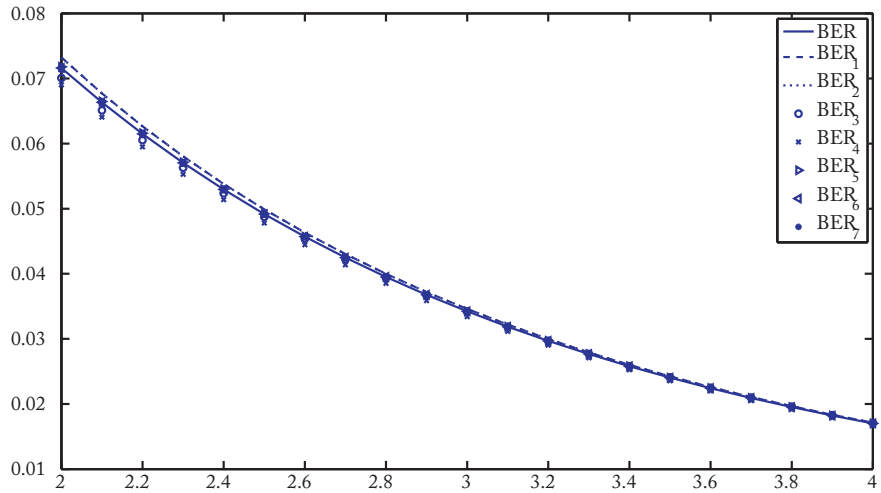}
\caption{Approximations for BER on $[2,4]$}\label{figure3}
\end{figure}

Now, in order to estimate the tightness of the new approximate expression and to compare with the tightness of the known
approximate expressions, for $i\in\{5,6,7\}$ let us consider the relative errors
$\varepsilon_{i}\triangleq\left(\mbox{BER}_i-\mbox{BER}\right)/\mbox{BER}.$ Table \ref{tableerror} contains
the relative errors $\varepsilon_i,$ where $i\in\{5,6,7\}$. Finally, in Fig. \ref{figure2} and \ref{figure3} the exact $\mbox{BER}$ and the approximate expressions $\mbox{BER}_i,$ where $i\in\{1,\dots,7\},$
are shown as functions of the SNR on the intervals $(0,2]$ and $[2,4].$ We find that our new approximate
expressions are quite accurate for the whole region of SNR.

\section{Conclusion}\label{section4}

In this correspondence we deduced a new tight upper bound for the BER of DQPSK with
 Gray coding over the AWGN channel
by using a new tight bound deduced for the Marcum $Q$-function. This
and the bounds from \cite{baricz} show that approximations of BER
for $\gamma=1$ obtained by using adaptive Simpson quadrature
(see \cite{weinberg}) and some previous approximations are not in the
interval determined by the lower and upper bounds.  Although the
estimations from \cite{cofe} and \cite{sun} are the best possible
estimations for large values of SNR, by using constants in the
convex combinations of the lower and upper bounds, they can be
improved by considering weight functions in the calculation of
approximations. The advantage of this technique is that we can
obtain better approximations also for small values of SNR. Based on
numerical experiments we proposed five new approximate BER
expressions for the AWGN channel, which use the best known bounds
and which are accurate in the whole region of SNR.

\newpage

\newpage

\begin{table}
\renewcommand{\arraystretch}{1.3}
\centering
\begin{tabular}
 {|c|c|c|c|c|} \hline
$\gamma$(dB) & BER & $\mbox{BER}_1$ & $\mbox{BER}_2$ & $\mbox{BER}_3$\\
\hline 1 & $1.639\times 10^{-1}$ & $1.739\times 10^{-1}$ & $1.731\times 10^{-1}$  &  $1.556\times 10^{-1}$ \\
\hline 2 & $7.161\times 10^{-2}$ & $7.324\times 10^{-1}$ & $7.322\times 10^{-2} $ & $ 7.007\times 10^{-2}$ \\
\hline 3 & $3.422\times 10^{-2}$ & $3.458\times 10^{-2}$ & $3.458\times 10^{-2}$ &  $3.416\times 10^{-2}$ \\
\hline 4 & $1.701\times 10^{-2}$ & $1.711\times 10^{-2}$ & $1.711\times 10^{-2}$  & $1.706\times 10^{-2}$ \\
\hline 5 & $8.648\times 10^{-3}$ & $8.683\times 10^{-3}$ & $8.683\times 10^{-3} $ & $8.677\times 10^{-3}$ \\
\hline 6 & $4.461\times 10^{-3}$ & $4.474\times 10^{-3}$ & $4.4745\times 10^{-3} $ &$ 4.473\times 10^{-3}$ \\
\hline 7 & $2.325\times 10^{-3}$ & $2.330\times 10^{-3}$ & $2.3308\times 10^{-3}$  &$ 2.33072\times 10^{-3}$ \\
\hline 8 & $1.221\times 10^{-3}$ & $1.224\times 10^{-3}$ & $1.22405\times 10^{-3}$  & $1.22404\times 10^{-3}$ \\
\hline 9 & $6.459\times 10^{-4}$ & $6.468\times 10^{-4}$ & $6.46883\times 10^{-4} $ &  $6.46881\times 10^{-4}$ \\
\hline 10 & $3.431\times 10^{-4}$ & $3.435\times 10^{-4}$ & $3.43588\times 10^{-4} $ & $3.43588\times 10^{-4}$ \\
\hline 11 & $1.830\times 10^{-4}$ & $1.832\times 10^{-4}$ & $1.83249\times 10^{-4} $ &  $1.83249\times 10^{-4}$ \\
\hline 12 & $9.798\times 10^{-5}$ & $9.807\times 10^{-5}$ & $9.80723\times 10^{-5}$  & $9.80723\times 10^{-5}$ \\
\hline
\end{tabular}
\caption{BER and the estimates $\mbox{BER}_1,$ $\mbox{BER}_2$ and $\mbox{BER}_3$}
\label{table1}
\end{table}

\begin{table}
\renewcommand{\arraystretch}{1.3}
\centering
\begin{tabular}
 {|c|c|c|c|c|} \hline
$\gamma$(dB) & $\mbox{BER}_4$ & $\mbox{BER}_5$ & $\mbox{BER}_6$ & $\mbox{BER}_7$\\
\hline 1 &$ 1.484 \times 10^{-1}$  & $ 1.677 \times 10^{-1}$ & $1.6383 \times 10^{-1}$&  $1.645 \times 10^{-1}$\\
\hline 2 &  $ 6.908 \times 10^{-2}$ & $ 7.179\times 10^{-2}$ & $7.162\times 10^{-2}$ & $ 7.133\times 10^{-2}$\\
\hline 3 &   $3.348 \times 10^{-2}$&  $3.423 \times 10^{-2}$ & $3.4226 \times 10^{-2}$&  $3.4233\times 10^{-2}$\\
\hline 4 &   $1.677 \times 10^{-2}$&  $1.7014 \times 10^{-2}$ & $1.7017 \times 10^{-2}$&  $1.7036\times 10^{-2}$\\
\hline 5 &   $8.5931 \times 10^{-3}$&  $8.6500 \times 10^{-3}$ & $8.6500 \times 10^{-3}$&  $8.6593\times 10^{-3}$\\
\hline 6 &   $4.4788 \times 10^{-3}$&  $4.4624 \times 10^{-3}$ & $4.4616 \times 10^{-3}$&  $4.4651\times 10^{-3}$\\
\hline 7 &   $2.3741 \times 10^{-3}$&  $2.3262 \times 10^{-3}$ & $2.3257 \times 10^{-3}$&  $2.3267\times 10^{-3}$\\
\hline 8 &   $1.2841 \times 10^{-3}$&  $1.2222 \times 10^{-3}$ & $1.2219 \times 10^{-3}$&  $1.2218\times 10^{-3}$\\
\hline 9 &   $7.1447 \times 10^{-4}$&  $6.4613 \times 10^{-4}$ & $6.4597 \times 10^{-4}$&  $6.4594\times 10^{-4}$\\
\hline 10 &  $ 4.1463\times 10^{-4}$ & $ 3.4326\times 10^{-4}$ & $3.4318\times 10^{-4}$ & $ 3.4317\times 10^{-4}$\\
\hline 11 &   $2.5591\times 10^{-4}$ &  $1.8311 \times 10^{-4}$ & $1.8307\times 10^{-4}$ & $ 1.8306\times 10^{-4}$\\
\hline 12 &  $ 1.7151\times 10^{-4}$ &  $9.8011 \times 10^{-5}$ & $9.7990\times 10^{-5}$ &  $9.7989\times 10^{-5}$\\

\hline
\end{tabular}
\caption{Estimates of BER by using $\mbox{BER}_4,$ $\mbox{BER}_5,$ $\mbox{BER}_6$ and $\mbox{BER}_7$}
\label{table2}
\end{table}

\begin{table}
\renewcommand{\arraystretch}{1.3}
\centering
\begin{tabular}
 {|c|c|c|c|} \hline
$\gamma$(dB) & $\varepsilon_5$ & $\varepsilon_6$ & $\varepsilon_7$\\

\hline 1 & $ 2.31\times 10^{-2}$ & $-4.51 \times 10^{-4}$ & $3.86\times 10^{-3}$\\
\hline 2 &   $2.52\times 10^{-3}$ & $ 1.96\times 10^{-4}$ & $-3.79\times 10^{-3}$\\
\hline 3 &   $1.17 \times 10^{-4}$ &$-1.36\times 10^{-5}$  & $1.94\times 10^{-4}$\\
\hline 4 &   $9.01 \times 10^{-5}$  &$2.66\times 10^{-4}$   &$1.38\times 10^{-3}$\\
\hline 5 &   $1.96 \times 10^{-4}$  &$1.93\times 10^{-4}$   &$1.26\times 10^{-3}$\\
\hline 6 &   $2.52 \times 10^{-4}$ & $8.90\times 10^{-5}$   &$8.65\times 10^{-4}$\\
\hline 7 &   $2.72 \times 10^{-4}$ & $4.55\times 10^{-5}$   &$5.04\times 10^{-4}$\\
\hline 8 &   $2.72 \times 10^{-4}$  &$2.54\times 10^{-5}$  &$-6.42\times 10^{-5}$\\
\hline 9 &   $2.62 \times 10^{-4}$  &$1.53\times 10^{-5}$  &$-3.32\times 10^{-5}$\\
\hline 10 &   $2.48\times 10^{-4}$  & $1.00\times 10^{-5}$ & $-1.66\times 10^{-5}$\\
\hline 11 &   $2.32   \times 10^{-4}$&$7.13\times 10^{-6}$  &$-6.71\times 10^{-6}$\\
\hline 12 &   $2.17  \times 10^{-4}$ &$5.44 \times 10^{-6}$ &$-3.87\times 10^{-7}$\\
\hline
\end{tabular}
\caption{Relative errors $\varepsilon_i$ of the approximate
expressions $\mbox{BER}_i,$ where $i\in\{5,6,7\}$.}
\label{tableerror}
\end{table}

\end{document}